# The emergence and evolution of the research fronts in HIV/AIDS research

David Fajardo-Ortiz*[1], Malaquias Lopez-Cervantes,[1] Luis Duran[1], Michel Dumontier,[2] Miguel Lara [3], Hector Ochoa,[4] and Victor M. Castano [5,*]

**Key worlds:** Complexity, normal science, scientific revolutions, network analysis

1. Facultad de Medicina, Universidad Nacional Autónoma de México, Mexico City, 04510, Mexico
2. Institute of Data Science, Maastricht University, Universiteitssingel 60, 6229 ER, Maastricht, The Netherlands.
3. Instituto de Biología, Universidad Nacional Autónoma de México, Mexico City, 04510, Mexico.
4. Colegio de la Frontera Sur, Chiapas, 29290, Mexico
5. Centro de Física Aplicada y Tecnología Avanzada, Universidad Nacional Autónoma de México, Queretaro, 76230, Mexico.

*Corresponding authors:
Victor M. Castano: vmcastano@unam.mx
David Fajardo-Ortiz: davguifaj@gmail.com

**Abstract**

In this paper, we have identified and analyzed the emergence, structure and dynamics of the paradigmatic research fronts that established the fundamentals of the biomedical knowledge on HIV/AIDS. A search of papers with the identifiers "HIV/AIDS", "Human Immunodeficiency Virus", "HIV-1" and "Acquired Immunodeficiency Syndrome" in the Web of Science (Thomson Reuters), was carried out. A citation network of those papers was constructed. Then, a sub-network of the papers with the highest number of inter-citations (with a minimal in-degree of 28) was selected to perform a combination of network clustering and text mining to identify the paradigmatic research fronts and analyze their dynamics. Thirteen research fronts were identified in this sub-network. The biggest and oldest front is related to the clinical knowledge on the disease in the patient. Nine of the fronts are related to the study of specific molecular structures and mechanisms and two of these fronts are related to the development of drugs. The rest of the fronts are related to the study of the disease at the cellular level. Interestingly, the emergence of these fronts occurred in successive "waves" over the time which suggest a transition in the paradigmatic focus. The emergence and evolution of the biomedical fronts in HIV/AIDS research is explained not just by the partition of the problem in elements and interactions leading to increasingly specialized communities, but also by changes in the technological context of this health problem and the dramatic changes in the epidemiological reality of HIV/AIDS that occurred between 1993 and 1995.

**Introduction**

The Human Immunodeficiency Virus/ Acquired Immunodeficiency Syndrome (HIV/AIDS) is a global health problem: over 70 million people have been infected with HIV, 35 million have died and 36.7 million people currently live with the disease **[1]**. HIV/AIDS is one of the most studied infection diseases with more than 260,000 papers (mentioning the topic) listed in GOPubMed **[2]** and more than 42,000 papers (mentioning HIV/AIDS in the title) in the Web of Science **[3]** spanning over thirty year of scientific research. HIV/AIDS is studied by a plurality of biomedical disciplines like epidemiology **[4],** virology **[5]**, immunology **[6]** or drug development **[7]** and non-biomedical disciplines like social sciences **[8]** and humanities **[9].** All the biomedical disciplines working on HIV/AIDS strongly rely on a solid scientific consensus, which explains the clinical manifestation of HIV/AIDS in terms of the virus interactions with the immune system cells; the behavior and demography of the immune system cells, and, most importantly, the virus interaction with the biomolecular machinery of the host cells **[10-12].** Two features are believed to be at the core of the scientific consensus on HIV/AIDS: the natural history of the HIV infection (the number of CD4+

cells and HIV RNA copies plotted over the time) **[11]** and the virus replication cycle (from the virus entry to the virus assembly, budding and maturation) **[10, 12]**.

Paradigms are the keystone of research communities **[13, 14]**, for they provide a foundation for members of the community; they also define the questions, the standards, the rules and the expected results that drive research efforts **[13, 14].** Paradigms of HIV/AIDS research are often presented in a timeline format **[12, 15].** However, while such a historical perspective is informative, they present two disadvantages: the first is that the selection of the most relevant discoveries is arbitrary, i.e., not supported by scientometric evidence, while the second disadvantage is that the paradigms are not presented as the key elements of the organizing process of the research communities.

The study of the emerging research fronts offers the possibility of analyzing the relationship between the paradigms and the organizational process of the scientific communities **[14, 16, 17]**. Research fronts can be considered as modules or clusters in a citation network of papers, i.e., sparse sub-networks of papers that exhibit dense connections **[18]**.

It must be pointed out that research fronts are the footprint of the scientific communities. That is, citation patterns of scientists exhibit homophily **[14,19],** which is caused by the scientists trend to cite those papers that focus in similar topics with a similar approach -and very often they cite those papers that strengthen the papers argumentation **[14, 19, 20]**. Citations tend to point toward those discoveries that the research (sub)communities consider the most relevant ones. i.e., the paradigms **[13, 14, 21-23].** Therefore, paradigms occupy the most central location in the citation networks; they are the seeds that organize the emergence of the research fronts **[14, 17].** To explain the emergence of the biomedical consensus on HIV/AIDS requires a study of the structure and dynamics of the research fronts.

Previous studies using the research fronts analysis approach were mainly focused in topics from engineering **[24, 25]**, biotechnologies **[18]** and scientometrics **[26]**. There are some studies that focused in the structure of the biomedical knowledge on specific diseases **[14, 16, 17, 27, 28]**. Our previous research has been particularly focused in the core region of the literature networks **[14, 17, 27]**. By doing this we have discovered the key feature of the organization of the knowledge on cervical cancer **[27]** and Ebola fever **[17].** Others have reported the evolution of research fronts in anthrax research **[16],** cancer research **[28]** and cardiovascular medicine **[28].** By analyzing the structure and evolution of HIV/AIDS knowledge we further our understanding of the nature of the biomedical knowledge discovery.

**Objective**

Through a combination of text mining and network analysis, we sought to understand the emergence and evolution of the research fronts (the footprints of the research communities) that produced the paradigmatic explanation of this disease.

**Methodology**

1. A search of papers on HIV/AIDS was performed in the Web of Science **[3]** during March, 2017. The search criteria were the following: TITLE: ("HIV AIDS") OR TITLE: ("Human immunodeficiency virus") OR TITLE: ("acquired immune deficiency syndrome") OR TITLE: (hiv-1). Refined by: DOCUMENT TYPES: ( ARTICLE ). Timespan: All years. Indexes: SCI-EXPANDED, SSCI, A&HCI, CPCI-S, CPCI-SSH, BKCI-S, BKCI-SSH, ESCI. 60,464 papers were found.
2. A network model was built with the papers found in the Web of Science by using the software HistCite **[29].** Then, the network model was analyzed and visualized with Cytoscape **[30].** The indegree distribution of the network was evaluated to determine if it fitted to a power law function ($y=ax^b$).
3. A core sub-network of papers with an indegree ≥ 28 was then closely examined. Normally, the indegree distribution in citation networks follows a power law function such that only a few papers are very well cited, while most papers are not **[31].** This applies to the case of HIV/AIDS research as we report. We selected the papers with an indegree ≥ 28 because they are a small and workable quantity of papers that account for nearly half of the communication process through the citations network as it is reported in the results section (The selected papers received 42,8911 out of 679,497 citations from the HIV/AIDS literature). Top cited papers appear related to the paradigmatic milestones of a particular research topic.
4. A cluster analysis based in the Newman modularity **[32]** was performed on the core sub-network using Clust&see, a Cystoscape plug-in **[33].** This analysis divided the sub-network of citation in several research fronts (clusters or modules of papers). This clusters are defined by Newman as "groups of vertices within which connections are dense but between which they are sparse" **[32].**
5. The sub-network was displayed by using the "yFiles organic" algorithm, which is based on

the force-directed layout **[30]**. This algorithm considers the nodes as charged particles that exhibit repulsive forces and the vertices as springs. In this layout, the papers that cite the same papers tend to stick together making easier the visualization of the research fronts.

6. The number of papers of each research front was plotted over the years in order to track the dynamics of the research fronts.
7. The content of the identified research fronts -the abstract of their papers- was analyzed with KH Coder **[34],** a software for quantitative content analysis (Text mining). KH Coder delivered several outputs. However, we considered that the most informative output was the list of the most distinctive words which provided key information about what was the main focus of the papers of each front. Additionally, the five papers with the highest indegree within each of the research fronts were identified. in order to provide a context to the reading of the text mining results.
8. Because front 1 "patient" is the largest and most central front according to our results, a cluster analysis was then performed on it by using Clust&see. The sub-modules that form front 1 were identified.

**Results**

**The network model**

60,464 published articles on HIV/AIDS were identified by keyword search over the Thomson Reuters Web of Science. 57,485 of these papers form a single network of 679,497 inter-citations. The structural network analysis performed by Cytoscape showed that the distribution of the indegree in this network fitted a power law function ($y=ax^b$, $a=51,954$, $b=-1.79$, correlation=0.827, R-squared=0.909). This means that a very small number of papers receive the overwhelming majority of citations while most papers receive few if any citations **[31, 35]**.

We selected papers with an indegree ≥ 28, that is, 5,933 documents. Together, these papers receive 63% of the inter-citations that form the whole network (42,8911 of 679,497), and would represent a relevant part of the historical core of the HIV/AIDS research as it was explained in the methodology section. These 5,933 highly cited papers formed a network of 86,963 inter-citations (Fig 1). The cluster network analysis identified fourteen clusters (or modules as defined by Newman). However, one cluster were too small to be considered relevant research fronts. The thirteen clusters were formed by 12,303, 9,115, 7,407, 6,746, 5,680, 4,763, 4,696, 3507, 2,861, 2,768, 2,597, 2,053, and 1,662 inter-citations.

**The organization and dynamics of the research fronts**

The main interaction among the seven research fronts is shown in Fig 2. In this figure, each of the research fronts is represented by a single node, and the edges represent the sum of the inter-citations among the fronts (Fig 2). Therefore, this graphical synthesis allows us to understand how the paradigmatic core of HIV/AIDS research was organized. However, the complete panorama can be only understood when Fig 2 is simultaneously read with Fig 3 and with Table 1 and S1 Table. Fig 3 shows the dynamics of the research fronts by plotting the number of citation papers per year of each research. In Fig 3, the fronts are shown in different plots according to the period of time in which the fronts reached their maximum number of papers per year: between 1990 and 1991 (Fig 3A), between 1996 and 1999 (Fig 3B), and between 2004 and 2007 (Fig 3C). S1 Table provides the detailed description of each research front, including its structural features;  the main topics of each research fronts according to the text mining analysis, and the list of the papers with the highest indegree within each research front. Table 1 groups the fronts according to the organization level (Individual, cellular-tissular and molecular) and the period in which the number of papers of each front peaked. The figures and the tables together offer an interesting view of the evolution and organization of HIV/AIDS research in the three first decades:

A peer reviewed and formatted version of this paper is available as: Fajardo-Ortiz D, Lopez-Cervantes M, Duran L, Dumontier M, Lara M, Ochoa H, et al. (2017) The emergence and evolution of the research fronts in HIV/AIDS research. PLoS ONE 12(5): e0178293.
https://doi.org/10.1371/journal.pone.0178293**Fig 1. The thirteen research fronts in the network model.** The model is displayed by using the "yFiles organic" algorithm. The color of nodes (representing the papers) and vertices indicates which research front they belong to.

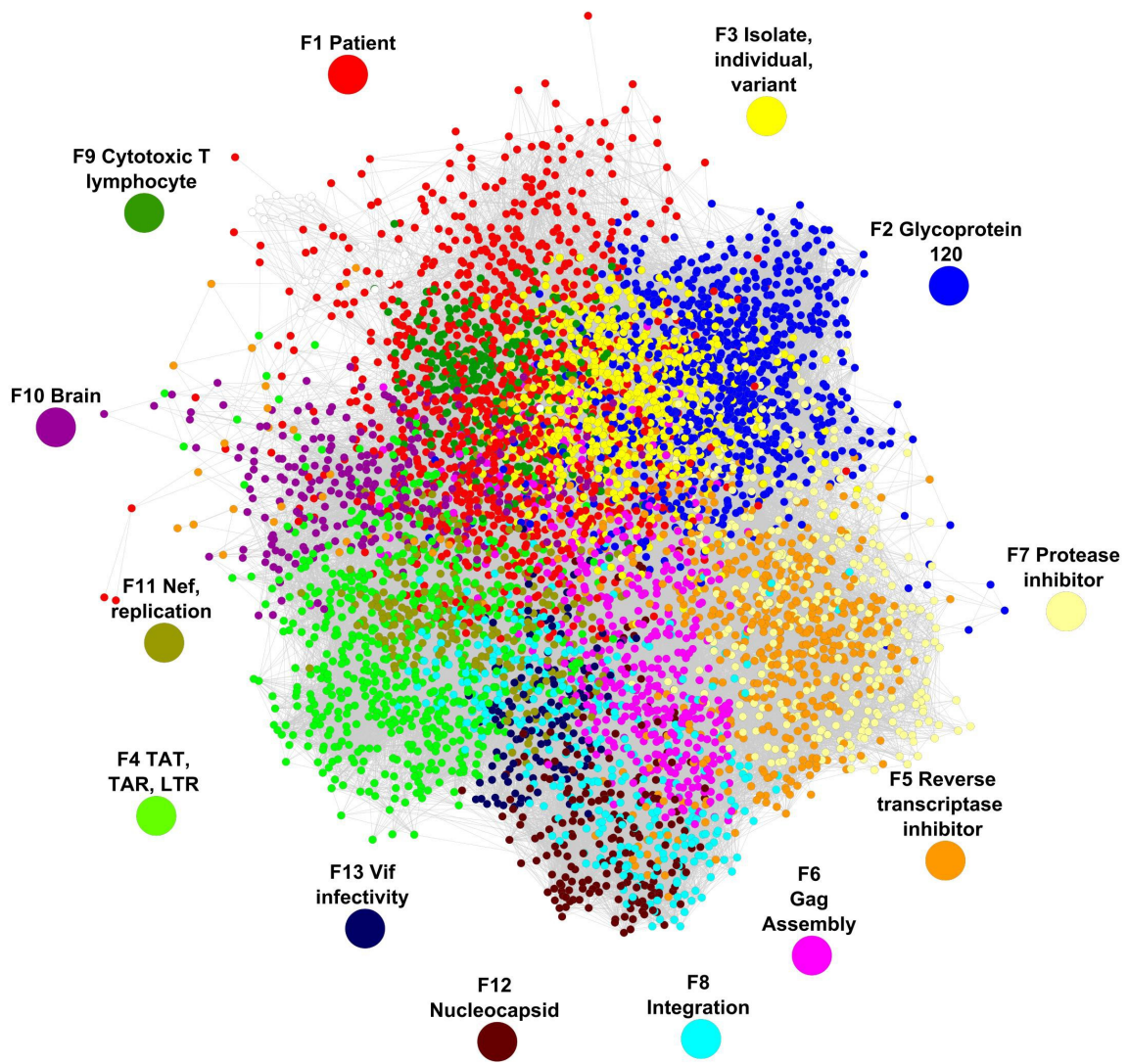

**Fig 2. Main interactions among the research fronts.** Each node represents one of the seven research fronts. The edges represent the sum of the inter-citations between two clusters. Only the interactions formed by a minimal of 500 inter-citations or the largest interaction (If the front have none interaction ≥ 500 inter-citations) of each front are shown.

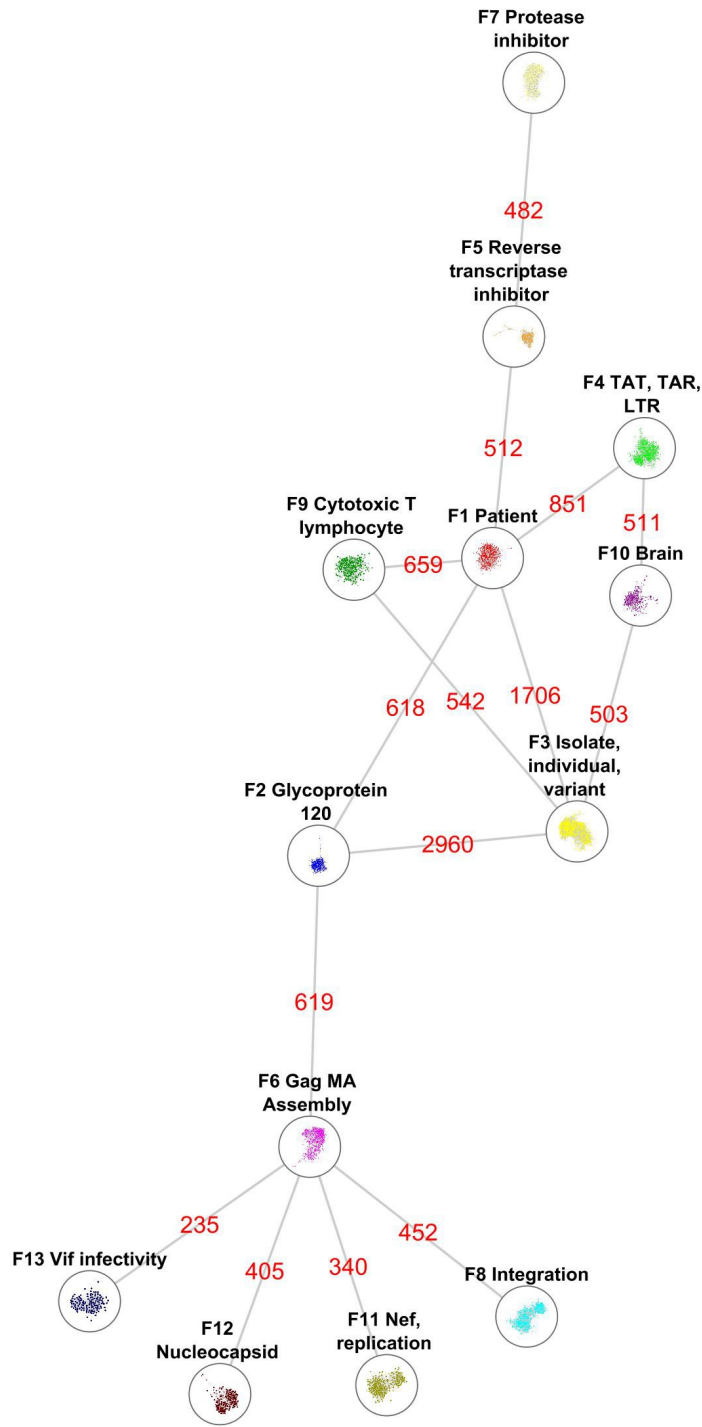

**Fig 3. Number of papers per year for each of the research fronts.** A: Research fronts whose number of papers peaked between 1990 and 1991, B: peaked between 1996 and 1999 and C: peaked between 2004 and 2007.

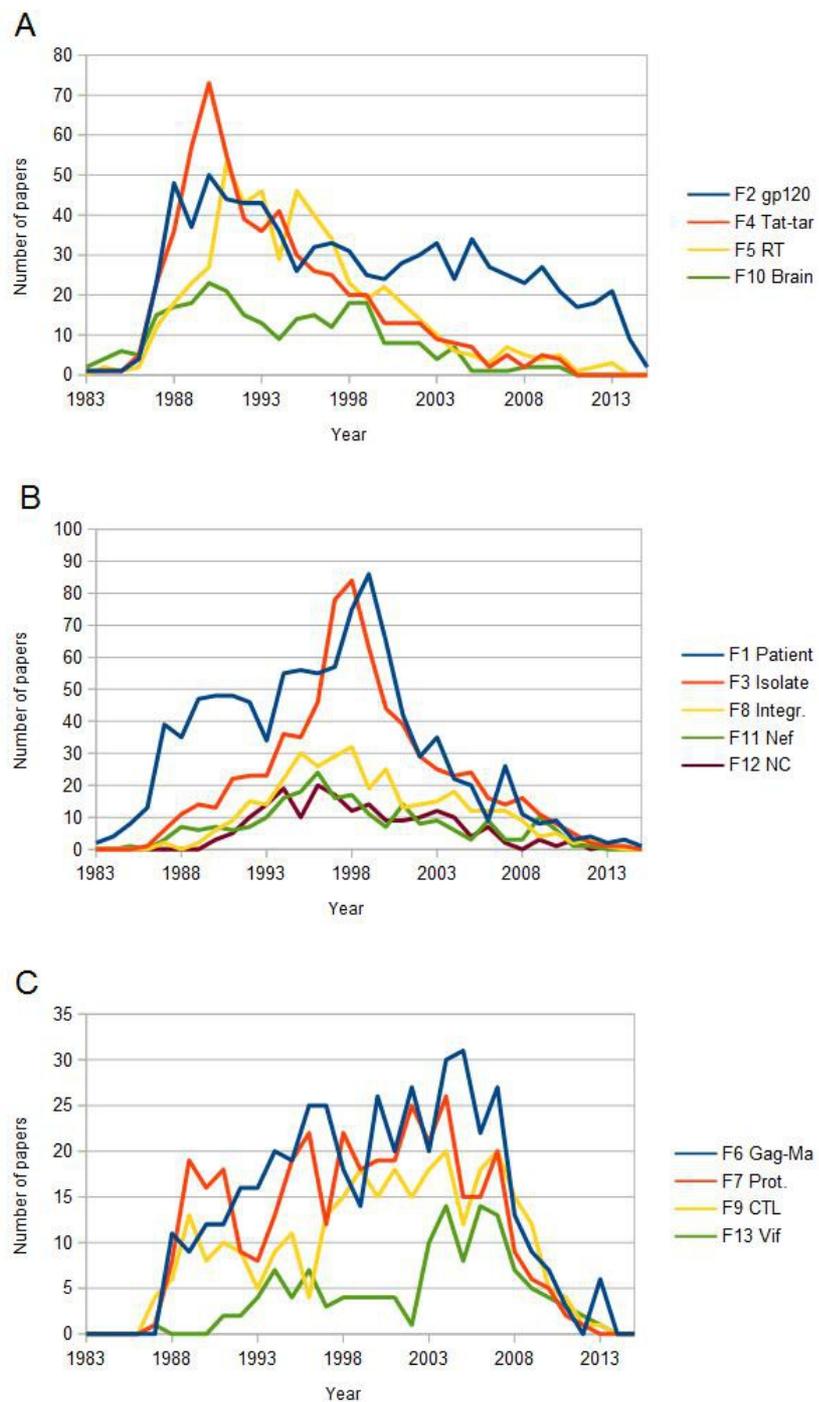

**Table 1.** Research fronts grouped by organization level (Individual, cellular-tissular and molecular) and by the period in which their number of papers peaked.

| Level/peak | 1990-1991 | 1996-1998 | 2004-2007 |
| --- | --- | --- | --- |
| Individual | | F1 Patient | |
| Cellular-tissular | F10 Brain | F3 isolate | F9 Cytotoxic T lymphocyte |
| Molecular | F2 gp120, F4 Tat-tar, F5 Reverse transcriptase | F8 Integrase, F11 Replication/Nef, F12 Nucleocapsid | F6 Gag/assembly, F7 protease, F13 Infectivity/Vif |

HIV/AIDS research is reductionistly-organized, following a hierarchy of biological structures and systems (Fig 2, Table 1). That is, at the core of the network is located the front 1 "patient" (Figs 1 and 2) which focus on the study of the HIV/AIDS phenomenon at individual-systemic level. Surrounding front 1, there are research fronts (Front 3 "isolate," front 9 "Cytotoxic T lymphocyte" and front 10 "brain" in Fgs 1 and 2) that are related to the study of specific events of the disease at cellular level (Table 1 and S1 Table ). In the most external part of the network model there are most of the fronts related to the study of molecular structures and mechanisms (Front 5 "reverse transcriptase inhibitor" front 6 "assembly," front 7 "protease inhibitor" front 8 "integration," front 11 "replication," front 12 "nucleocapsid" front and 13 "infectivity"; see Figs 1 and 2 and Table 1 and S1 Table). On the other hand, Front 2 "glycoprotein 120" and front 5 "reverse transcriptase inhibitor" are strongly connected to front 1 "patient" (Figs 1 and 2) However, the research in these fronts (2 and 5) is oriented to the development of treatments (immuno-therapies and small molecules drugs, respectively), which may explain their strong connection to front 1. Notice that fronts 3 and 2 function as transition zones in HIV/AIDS research connecting the different levels of observation (systemic, tissular-to-cellular and molecular) (Figs 1 and 2).

On the other hand, the research fronts clearly can be grouped in three different periods of time in which the fronts reach their maximum number of papers per year : 1990-1991, 1996-1999 and 2004-2007 (Fig 3). In order to properly read Fig 3 it is important to keep in mind the dramatic changes in the epidemiology of HIV/AIDS in the United States (USA) that happened between 1993 and 1995 **[36]**. In that period, the number of AIDS diagnosis and deaths reached their maximum and then declined**[36]**. Simultaneously, in 1995, the number of persons living with HIV began to rise**[36]**. Therefore, we can consider the existence of two stages in the history of HIV/AIDS: before 1995 in which AIDS was the main concern and after 1995 when HIV infection is at the center of HIV/AIDS research. A second important consideration to understand Fig 3 is that the phase of expansion or growth in science (the "normal" science of Thomas Kuhn) follows the publication of those scientific achievements that organize the subsequent research[13]. This would explain that the peaks in Fig 3 generally occurred years after the publication of the papers with the highest degree (Fig 3 and S1 Table). The peaks in figure can be considered a delayed response to fundamental events and discoveries in the history of HIV/AIDS research. A third consideration is that the network model is made from the ten percent of papers with the highest indegree. Therefore, the succession of research fronts observed in Fig 3 does not mean the end of the research on specific topics but that these topics are not longer in the core of HIV/AIDS research.

Fronts 2 "glycoprotein 120," 4 "tat-tar," 5 "reverse transcriptase inhibitor" and 10 "brain" emerged immediately after front 1 and peaked in the 1990 and 1991 years. The expansion of these fronts in this early stage in the history of HIV/AIDS research suggests that these fronts are relevant to the description, explanation or intervention of AIDS. For example, it has been pointed out that tat (Trans-activator of transcription) protein, which is essential for virus replication, could be involved in the progression to AIDS and in the development of Kaposi's sarcoma lesions.[37, 38] Along the same line, the interaction between glycoprotein 120 and CD4 is the first event in the replication cycle and is considered fundamental to virus entry.[39] It is important to keep in mind that the depletion of lymphocytes expressing CD4 is considered the most severe hematological feature of AIDS.[39] Similarly, encephalopathy is one of the most dominant feature of AIDS.[40] Finally, a reverse transcriptase inhibitor, zidovudine (AZT) was the first drug approved by the United States Food and Drug Administration (FDA) to treat AIDS.[41]

Research fronts 1 "patient," and 3 "isolate," reached their maximum number of papers per year between 1996 and 1999 (Fig 3). The peaks of these fronts follow the changes in the epidemiology of HIV/AIDS in the USA. Therefore, these fronts are possibly related to a collective response from the scientific community to the new reality of the disease. The research in front 1 is the largest, central and most clinical among the fronts (Figs 1 and 2, and S1 Table ). This front connects the clinical and epidemiological manifestations of HIV/AIDS with their explanation at a cellular level. Because of the size, the centrality and clinical relevance, we decided to perform a second round of cluster analysis to identify the sub-modules that may conform front 1. We plotted the contribution of each sub-module to the evolution of front 1 in Fig 4. Sub-module 1A, 1D and 1E are the key components of the 1999's peak (Fig 4). The papers with the highest indegree in sub-modules 1A, 1D and 1E are, respectively, "Reduction of maternal-infant transmission of human immunodeficiency virus type 1 with zidovudine treatment,"[42] "Rapid turnover of plasma virions and CD4 lymphocytes in HIV-1 infection"[43] and "Identification of a reservoir for HIV-1 in patients on highly active antiretroviral therapy".[44] These papers report and explain fundamental changes in the clinical reality of HIV/AIDS produced by the implementation of anti-retroviral therapies. Front 3 "isolate," to the study of HIV tropism, i.e., the differential capacity of the HIV strains to infect and replicates in different cell types (Table 1). Importantly, the availability of screening tools that allowed the identification of asymptomatic individuals infected with HIV and the use of anti-retroviral therapies make extensively available the blood and tissue samples from the patients that were fundamental to the emergence of front 3.

**Fig 4. The contribution of the sub-modules to the evolution of the research front 1.**

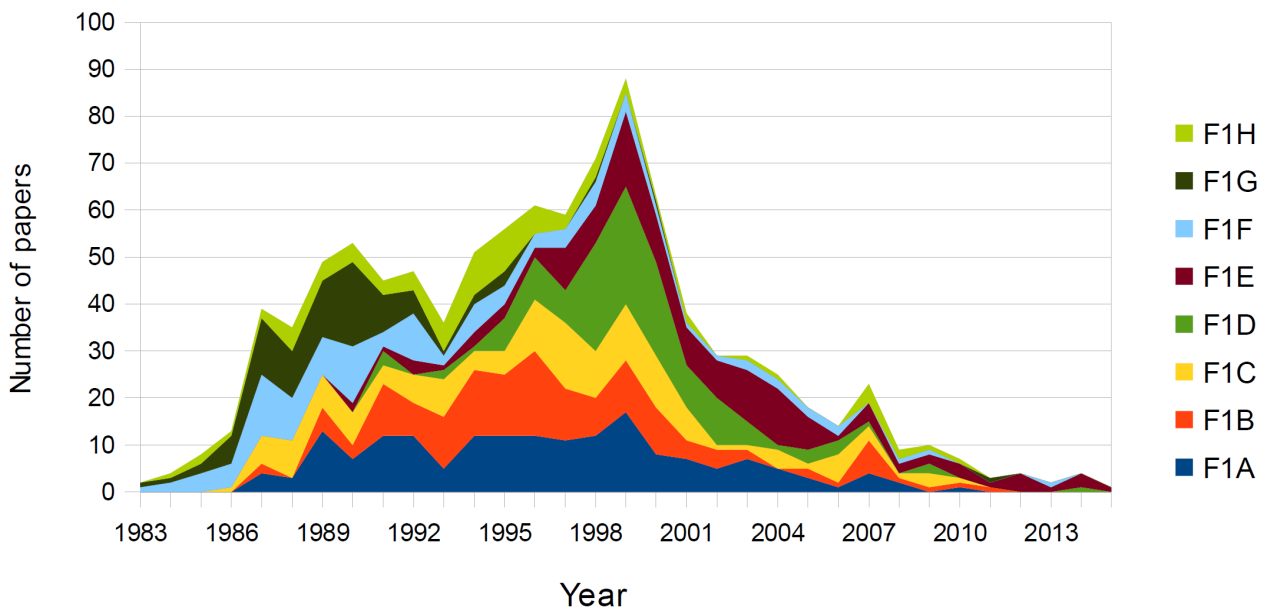

On the other hand, most of the research fronts specialized in the study of specific molecular mechanisms and structures (fronts 6 to 9 and fronts 11 to 13) peaked either in the 1996-1999 or 2004-2007 periods (Fig 3). It is important to notice that all these fronts emerged at the end of the first decade of HIV/AIDS research. The difference between fronts peaking in the second and the third periods is that the former decline earlier. In order to understand the evolution of these fronts is important to keep in mind that the scientific specialization is a continuous process of solving problems that follows the establishment of a paradigm (HIV-1 as the etiological agent of HIV/AIDS). [13] In that sense, the decline of fronts 8, 11 and 12 (Fig 3B) may be so because the scientific problem is either essentially solved or the changes in the HIV/AIDS epidemiology made less relevant the topics related to these fronts.

**Discussion**

According to the Cytoscape analysis, the network of HIV/AIDS papers displays a power law distribution of their citations, which has important methodological implications. The first implication is that a research front (a citation network module) could be formed by other research fronts, which in turn can be partioned into sub-modules **[44]**. The second implication is that the nodes (papers) with the highest indegree tend to be more "cosmopolitan" i.e., they have the lowest clustering coefficient values **[44]**. That is, they could belong simultaneously to several fronts or any of them. Therefore, there are not clearly defined frontiers dividing the research fronts. However, the standard in the scientometrics study of research fronts seems to be to use clustering methods that define frontiers between the modules **[14, 17, 18, 24, 25, 27, 28]**, probably because this is make much more understandable the community structure in a literature network. Moreover, our analysis reveals the front that is most relevant to each paper. Finally, the most important implication is that in a hierarchical literature network the most the papers with the highest indegree are related with the paradigms that organize a research field or topic **[14]**. Top cited papers have been extensively used to identify the scientific achievements that establish the standards of research practice of a particular community **[14, 17, 22, 23, 27, 28, 45, 46]**. There are no set guidelines on the proportion of top cited papers that should be selected. However, there is a trade off between selecting the most informative papers and maintaining diversity of the information **[14, 17]**. In this work, we used a minimal indegree of 30 to select the top cited papers, and in turn obtaining a considerable percentage of the citations. The selected papers consist of only ten percent of the network but they effectively account for two thirds of the citations. The selected papers are a reasonable representation of the paradigmatic core of HIV/AIDS research.

Once the HIV was recognized as the etiological factor HIV/AIDS research entered in normal science mode that is characterized by a high productivity and for a specialization of the researchers. By specialization we refer to "concentrate exclusively upon the subtlest and most esoteric aspects of the natural phenomena that concern his group" **[12]**. Once the paradigms are established, researchers focus on the details, the smaller range problems and solutions that the current paradigm provide **[13]**. Our results suggest that the emergence of several of the specialized research fronts was caused by the partition of the general problem in interacting elements. That is, HIV/AIDS research could be understood to some extent as a particular instance of part-whole science in which paradigms determine the abstraction of the parts that are considered the most relevant to explain the whole phenomenon **[47, 48]**.

The general structure and evolution of the research fronts in HIV/AIDS research shares similarities to that of anthrax and Ebola. The evolution of anthrax investigation began with a preliminary on the immunology of the disease **[16]**. From this, four research fronts emerged: "anthrax gene sequencing", "vaccine research", secondary research on PA (protective antigen) and LF ( lethal factor), and "making and purifiying toxin" **[16]**. Subsequently, the research front on PA and LF split in three fronts: "specific PA research", PA mediated delivery of other substances" and specific "LF research" **[16]**. Similarly, the evolution of the fronts in Ebola research are marked by a front related to the report of the epidemiology and the clinical manifestation of the disease **[17]**. A second front provide an explanation of the disease at tissue-cellular level **[17]**. Then, research on Ebola split into four research fronts, each one specialized in one different virus protein **[17]**. There is also a front aimed to the development of vaccines and other immunotherapies **[17]**. Similarly, the emergence of the fronts in HIV/AIDS research started with a general research front that provided the pathology of the disease and subsequently split into specialized fronts focused on the study of specific molecular mechanism of the virus replication cycle. In all three cases, the specialization of the research led to the emergence of research fronts focused in the study of the parts that are thought to be key in explaining the diseases. A report on the emergence of the research fronts in cancer and cardiovascular diseases showed that the specialization process in these types of diseases is complex **[28]**. Jones et al. reported fronts specialized on microarrays, targeted therapies, clinical trials, epidemiology and molecular etiology in cancer research **[28],** while in cardiovascular diseases the fronts are organized around drug-eluting stents, anti-platelet agents, pacemakers, hypertension and atrial fibrillation **[28]**. The difference between these two groups of diseases is that HIV/AIDS, anthrax and Ebola are infectious diseases with a clearly identified etiological agent while cancer and cardiovascular diseases are both complex multifactorial diseases **[49]**.

This is the first time that the complex organization (and the evolution) of HIV/AIDS research is reported. Our research provides fundamental knowledge concerning the emergence of the paradigmatic explanation for HIV/AIDS and therefore makes a contribution to the understanding of the nature of biomedical knowledge. In addition, our work suggests that the development of the paradigmatic knowledge on HIV/AIDS in terms of the emergence and evolution of the research fronts followed two different routes. First, the emergence of the specialized fronts (molecular mechanism and structures and cellular process) was caused by the division of the general problem in their key process, element and interactions, which is related to the concept of part-whole science. Second, the dynamics of the fronts, particularly the evolution of front 1 "patient" and 2 "isolate", appears to represent an adaptive and collective response from the scientific community to changes in the epidemiological (the decline in the morbidity and mortality of AIDS in the USA) and technological (the availability of treatments and screening tools) context of this health problem.

**Supporting information**

**S1 Fig. Description of the research fronts.** Structural properties; top 10 distinctive worlds, and list of the five papers with the highest indegree within each of the research fronts. See: http://journals.plos.org/plosone/article/file?type=supplementary&id=info:doi/10.1371/journal.pone.0178293.s001

**Author contributions statement**

All authors contributed to the interpretation of results and writing of the paper. DF-O and VMC designed the research. DF-O performed the structural, dynamics, and content analyses in the literature network on HIV/AIDS. All authors read and approved the final manuscript.

**Competing interests**

The authors declare no competing financial interests.